\documentclass[amssymb,amsmath,prl,twocolumn,floatfix,showpacs,superscriptaddress]{revtex4}
\usepackage{graphicx}
\begin{document}
\title{Coherent electronic transfer in quantum dot systems using adiabatic passage.}

\author{Andrew D.~Greentree}
\affiliation{Centre for Quantum Computer Technology, School of
Physics, The University of Melbourne, Melbourne, Victoria 3010,
Australia.}
\affiliation{Centre for Quantum Computer Technology, School of
Physics, The University of New South Wales, Sydney, NSW 2052,
Australia.}

\author{Jared H.~Cole}
\affiliation{Centre for Quantum Computer Technology, School of
Physics, The University of Melbourne, Melbourne, Victoria 3010,
Australia.}

\author{A.~R.~Hamilton}
\affiliation{Centre for Quantum Computer Technology, School of
Physics, The University of New South Wales, Sydney, NSW 2052,
Australia.}

\author{Lloyd C.~L.~Hollenberg}
\affiliation{Centre for Quantum Computer Technology, School of
Physics, The University of Melbourne, Melbourne, Victoria 3010,
Australia.}

\date{September 17, 2004}

\begin{abstract}
We describe a scheme for using an all-electrical, rapid, adiabatic population
transfer between two spatially separated dots in a triple-quantum dot system.
The electron spends no time in the middle dot and does not change its energy during the transfer process. Although a coherent population
transfer method, this scheme may well prove useful in incoherent
electronic computation (for example quantum-dot cellular automata)
where it may provide a coherent advantage to an otherwise
incoherent device.  It can also be thought of as a limiting case of type II quantum computing, where sufficient coherence exists for a single gate operation, but not for the preservation of superpositions after the operation.  We extend our analysis to the case of many intervening dots and address the issue of transporting quantum information through a multi-dot system.
\end{abstract}


\maketitle

%%%%%%%%%%%%%%%%%%%%%%%%%%%%%%%%%%%%%%%%%%%%%%%%%%%%%%%%%%%%%%%%

\section{Introduction}
Practical devices based on coherent quantum mechanical properties
are widely seen as being one of the next major advances in electronics.  This search is typified by the keen interest in quantum
computation \cite{bib:NielsenBook}.  In this paper, however, we
seek first to explore a slightly different aspect of coherent quantum
devices, by investigating how coherence can be used to enhance the
capabilities of an otherwise incoherent electronic device. We will then consider extensions  where quantum information, rather than classical information, is transported around a quantum dot system.

We consider how to utilize coherent charge transfer in an
incoherent device, i.e. one where the initial and final states of
the device are eigenstates of the system Hamiltonian, but intermediate states are formed by
coherent superposition states.  Applications which might benefit
by such coherent charge transfer could be single electron pumps
\cite{bib:GrabertNATO1992} or electronic digital logic (for
example through quantum-dot cellular automata (QDCA) \cite{bib:QDCA}),
but our ideas are general and should have wide applicability. By
exploring this regime we seek to increase our understanding of
quantum coherence and exploit coherent effects before the
technical difficulties of building truly global coherent quantum
devices have been solved.  As such one may even consider our proposal as realising a limiting case of a type-II quantum computer \cite{bib:TypeII} where there is enough coherence to effect a single gate operation (a swap gate) but not enough coherence to maintain any superpositions after the initial transfer is completed.

Mechanisms of coherent population transfer can be broadly classified
into two classes, nonadiabatic and adiabatic. Nonadiabatic methods
are usually typified by control of tunnelling to realize $\pi$ or
similar pulses, which transfer population from one state to another
coherently and rapidly.  Nonadiabatic processes require exacting
control over pulse areas and tunnelling rates that are fast compared
with all relevant dephasing times.  Although nonadiabatic population
transfer is a goal, the sensitivity to noise of nonadiabatic
transfer can place prohibitive requirements on pulse switching times
in certain implementations.  Adiabatic transfer is usually typified
by quasi-static population transfer.  Such methods are usually slow
and are accompanied by high fidelity and relative insensitivity to
gate errors and other external noise.  Between these regimes lie
adiabatic passage techniques which combine the advantages of
robustness from adiabatic techniques with well-defined trajectories
to realize population transfer rates approaching those of the
nonadiabatic regime.  We differentiate adiabatic passage from other
adiabatic transfer processes by the \textit{simultaneous} modulation
of at least two system parameters to realize a desired trajectory
through the Hilbert space.  For a further discussion of adiabatic
and nonadiabatic timescales in quantum electronic transfer see
Ref.~\onlinecite{bib:GreentreeAT2003}.

Adiabatic passage techniqes were originally developed for NMR and optical applications.  For a good review see Ref.~\cite{bib:VitanovARPC2001}.  There have been several investigations of the applications of electromagnetically mediated adiabatic passage in solid-state quantum computing, see for example Ref.~\cite{bib:EMAP}.  There are
reasonably few proposals for adiabatic passage without using
electro-magnetic radiation in the literature, although Brandes and
Vorrath consider adiabatic passage through a double-dot system
\cite{bib:BrandesPRB2002}; Barrett and Milburn considered using adiabatic passage as a method for determining dephasing rates in a qubit \cite{bib:BarrettPRB2003}; and more recently Zhang \textit{et al.} \cite{bib:ZhangPRA2004} have proposed a scheme for generating entanglement in a triple-dot, two-electron system via adiabatic passage.  Adiabatic passage has been applied to the problems of reading out a charge-superposition qubit \cite{bib:GreentreeAT2003} and spin qubit \cite{bib:GreentreeSpin}.

We propose adiabatic passage through a triple-well system.  We term this method CTAP (Coherent Tunnelling by Adiabatic Passage) and it constitutes an all-electrical analogue to the more well-known  optical STIRAP (Stimulated Raman Adiabatic Passage) scheme for population transfer in a three-level
$\Lambda$ system.  Independantly, Eckert \textit{et al.} \cite{bib:EckertPRA2004} have considered the realization of an analogous STIRAP style scheme with a neutral atom in a triple well potential formed in an optical trap. Such schemes are more complicated than two-state adiabatic fast passage, but have the advantage
that the electron energy is unchanged throughout the population
transfer.  This consideration is important when considering the
trade-off between heat dissipation and gate speed in QDCA
\cite{bib:TimlerJAP2003}, and our scheme could be used to realise true adiabatic clocking of a QDCA, as opposed to the quasi-adiabatic clocking schemes reported recently \cite{bib:TothJAP1999}.  Maintaining constant energy is also important for preventing an extra dynamical phase appearing when quantum information is being transported, as will be discussed below.

\section{Transport in a triple-dot system}

\begin{figure}[tb!]
\includegraphics[width=5.5cm,clip]{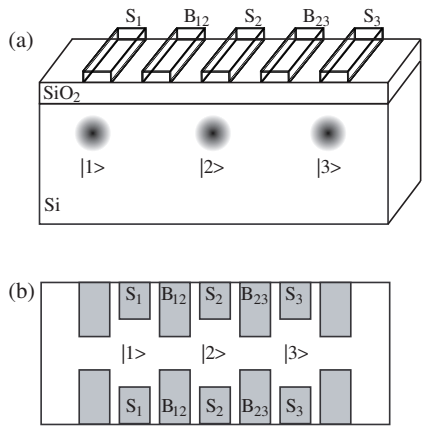}
\caption{\label{fig:TripleWell} (a) Schematic of the triple donor
system realized with individually placed donors in silicon.
Two of the donors are assumed ionized, the other neutral.
The donors are labelled from left to right $1,2,3$.
The position eigenstates of the electron are $|1\rangle$, $|2\rangle$, $|3\rangle$, corresponding to the donor labels, with energy ${\cal E}_{\alpha}$ for
$\alpha=1,2,3$ which can be controlled by the $S$ gates.
Coherent tunnelling is between nearest neighbors
only, with rates $\Omega_{12}$ and $\Omega_{23}$.  These
rates are controlled by external $B$ gates and can
therefore be time varying. (b) Another possible implementation
in a 2-D electron gas.  Light gray rectangles
represent the $S$ and $B$ gates. The gates are set so that only one electron
is allowed in the system, with only one electronic state per region.  There are electron reservoirs to the left and right hand side.  Readout for both schemes is assumed via sensitive electrometers (not shown).}
\end{figure}

A schematic of our system appears in Fig.~\ref{fig:TripleWell}.  We have a
triple-well system $|1\rangle$, $|2\rangle$, $|3\rangle$, where we
wish to achieve coherent population transfer from $|1\rangle$ to
$|3\rangle$ without any population being in state $|2\rangle$.  The system is controlled by shift gates, $S_i$ which control the energy of dot $i$, and barrier gates, $B_{i,i+1}$ which control the tunnelling rate between dots $i$ and $i+1$.  Our scheme is applicable to any system where the coherent tunnelling times can be varied from much slower than the T$_1$ time (population relaxation) to several times faster than the T$_2$ time (coherence relaxation).  An ideal system to consider for implementing this proposal would be the the phosphorus in silicon donor scheme introduced in \cite{bib:Hollenberg} for
quantum computing, however it could equally
well be realized in a superconducting system \cite{bib:NakamuraNat1999}, two-dimensional electron gas \cite{bib:2DEG}, or even conceivably in a metallic triple-dot structure realised via small buried phosphorus clusters \cite{bib:PClusters}\footnote{In a metallic structure there will be a very
large dephasing rate, however because electronic transfer is
`one-way' in this scheme, the problem of electron registration
associated with observing coherent oscillations is avoided, hence
coherent transfer should be observable even without Rabi-like
oscillations.}. To proceed, we write down the Hamiltonian for the triple-well system in matrix form, with state ordering $|1\rangle$, $|2\rangle$, $|3\rangle$
\begin{eqnarray}
{\cal H} &=& \hbar \left[%
\begin{array}{ccc}
  0 &  -\Omega_{12} & 0 \\
  -\Omega_{12} & \Delta/\hbar & -\Omega_{23} \\
  0 & -\Omega_{23} & 0 \\
\end{array}%
\right]. \label{eq:Ham}
\end{eqnarray}
where $\Omega_{\alpha\beta}=\Omega_{\alpha\beta}(t)$ is the coherent tunnelling rate between position eigenstates
$|\alpha\rangle$ and $|\beta\rangle$ and $\Delta=E_2-E_1=E_2-E_3$.
In practice, to maintain the state energies will require
the use of $S$ gates for tuning and compensation.  The eigenstates
of the Hamiltonian of Eq.~\ref{eq:Ham} are \cite{bib:VitanovARPC2001}
\begin{eqnarray}
|{\cal D}_+\rangle &=& \sin \Theta_1 \sin \Theta_2 |1\rangle +
        \cos \Theta_2 |2\rangle +
        \cos \Theta_1 \sin \Theta_2 |3\rangle, \nonumber \\
|{\cal D}_-\rangle &=& \sin \Theta_1 \cos\Theta_2 |1\rangle -
        \sin \Theta_2 |2\rangle +
        \cos \Theta_1 \cos \Theta_2|3\rangle, \nonumber \\
|{\cal D}_0\rangle &=& \cos \Theta_1|1\rangle +
         0 |2\rangle -
        \sin \Theta_1 |3\rangle,
\label{eq:DressedStates}
\end{eqnarray}
where we have introduced
\begin{eqnarray}
\Theta_1 &=& \arctan \left(\Omega_{12}/\Omega_{23} \right), \nonumber
\\
\Theta_2 &=& \frac{1}{2} \arctan
\left[\left(\sqrt{(2\hbar\Omega_{12})^2 + (2\hbar\Omega_{23})^2}\right) / \Delta
\right].
\end{eqnarray}
The energies of these states are:
\begin{eqnarray}
{\cal E}_\pm &=& \frac{\Delta}{2} \pm
\frac{1}{2}\sqrt{(2\hbar\Omega_{12})^2
+ (2\hbar\Omega_{23})^2 + \Delta^2}, \nonumber \\
{\cal E}_0 &=& 0. \label{eq:Eigenenergies}
\end{eqnarray}

Our aim is to induce population transfer from state $|1\rangle$ to
$|3\rangle$ by maintaining the system in state $|{\cal
D}_0\rangle$ and changing the characteristics of $|{\cal
D}_0\rangle$ from $|1\rangle$ at $t=0$ to $|3\rangle$ at
$t=t_{\max}$ by appropriate control of the tunnelling rates,
without population leakage into the other eigenstates.  Provided we start in an eigenstate, and transform our Hamiltonian adiabatically, we will remain in the same eigenstate.  The criterion for adiabaticity is \cite{bib:VitanovARPC2001}
\begin{eqnarray}
|{\cal E}_0 - {\cal E}_{\pm}| \gg |\langle \dot{\cal D}_0|{\cal
D}_{\pm}\rangle|.
\end{eqnarray}

\begin{figure}[tb!]
\includegraphics[width=5.5cm,clip]{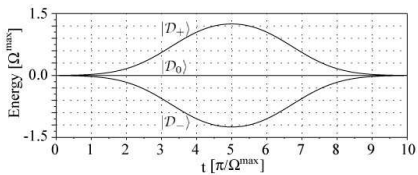}
\caption{\label{fig:3sEnergy} Energies of the states $|\mathcal{D}_+\rangle$, $|\mathcal{D}_0\rangle$ and $|\mathcal{D}_-\rangle$ from top to bottom, respectively, through the CTAP pulse sequence described in Eq.~\ref{Eq:CTAPPulses} with $\Delta=0$.}
\end{figure}

It is usual to apply Gaussian pulses to realize the
adiabatic transfer, and we consider pulses of the
form
\begin{eqnarray}
\Omega_{\alpha\beta} = \Omega_{\alpha\beta}^{\max}
    \exp\left[-(t-t_{\alpha\beta})^2/(2\sigma^2_{\alpha\beta})\right],
\end{eqnarray}
where $t_{\alpha\beta}$ and $\sigma_{\alpha\beta}$ are the peak
time and standard deviation of the control pulse modulating the tunnelling
rate between states $|\alpha\rangle$ and $|\beta\rangle$.  For
simplicity we set the maximum tunnelling rates and standard deviations for each transition to be equal,
i.e. $\Omega_{\alpha\beta}^{\max} = \Omega^{\max}$ and
$\sigma_{\alpha\beta}=\sigma$, and set $\Delta=0$ in the discussion that follows.  Under these conditions, transfer is optimized when the width of the pulses equals the time delay between the pulses \cite{bib:GaubatzJChemPhys1990}, with total pulse time  $t_{\max}$, we choose $\sigma = t_{\max}/8$ and the barrier controlled tunelling rates
\begin{eqnarray}
\Omega_{12}(t) = \Omega^{\max} \exp
    \left[ -\left( t-\frac{t_{\max}+\sigma}{2} \right)^2
    /(2 \sigma^2) \right], \nonumber \\
\Omega_{23}(t) = \Omega^{\max} \exp
    \left[ - \left(t-\frac{t_{\max}-\sigma}{2} \right)^2/(2 \sigma^2) \right].
\label{Eq:CTAPPulses}
\end{eqnarray}
This ordering, where $\Omega_{23}$ is applied \textit{before} $\Omega_{12}$ is known as the counter-intuitive pulse sequence and has significant advantages in improving transfer fidelity over other pulse sequences.  Under these conditions, the energies of the eigenstates are shown in Fig.~\ref{fig:3sEnergy} for $t_{\max} = 10 \pi /\Omega$.

In order to proceed, we numerically solve the master equations for the density matrix, $\rho$,
\begin{eqnarray}
\dot{\rho} = -\frac{i}{\hbar} \left[\mathcal{H},\rho\right] + \Gamma \left[\rho-\mathrm{diag}(\rho)\right], \label{eq:DensMatClosed}
\end{eqnarray}
where $\Gamma$ is the $T_2$ (pure dephasing) rate, assumed to act equally on all coherences.  As we are primarily considering solid-state systems, we ignore the $T_1$ rates, as $T_2$ is generally much faster, although such rates could be added in principle.

\begin{figure}[tb!]
\includegraphics[width=5.5cm,clip]{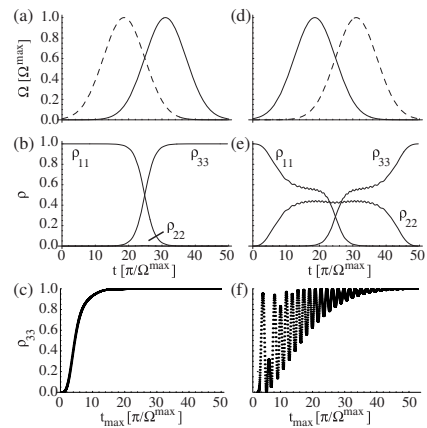}
\caption{\label{fig:AdiabaticTransfer} Charge transfer using both counter-intuitive [Fig.~\ref{fig:AdiabaticTransfer}(a) - \ref{fig:AdiabaticTransfer}(c)] and intuitive [Fig.~\ref{fig:AdiabaticTransfer}(d) - \ref{fig:AdiabaticTransfer}(f)] pulse sequences with no dephasing ($\Gamma=0$). (a) Tunnelling rates as a function of time (in units of $\pi/\Omega^{\max}$, $\Omega_{12}$ is the solid line, $\Omega_{23}$ is the dashed line. (b) Populations as a function of time as the trajectory of (a) is followed.  Note the smooth migration of population from $|1\rangle$ to $|3\rangle$ with no population in $|2\rangle$. (c) Final populations as a function of $t_{\max}$ (in units of $\pi/\Omega^{\max}$) where the pulse sequence and pulse widths are determined by Eq.~\ref{Eq:CTAPPulses}. (d) Intuitive pulse sequence, as before $\Omega_{12}$ is the solid line and $\Omega_{23}$ is the dashed line. (e) Populations as a function of time along the trajectory of (d).  In this case there are significant oscillations and a relatively large population in $|2\rangle$, this pulse ordering is clearly unsatisfactory for high fidelity transport. (f) as (c) but for the intuitive pulse ordering.  Note that in this case, to get high-fidelity transfer, more time is required.}
\end{figure}

In Fig.~\ref{fig:AdiabaticTransfer} we present results showing the population transfer using the counter-intuitive and intuitive pulse orderings without dephasing.  As long as the adiabaticity criteria
is satisfied, the fidelity of population transfer is very high.  In
practise the maximum possible transfer rates will be a few times greater than $\Omega^{\max}$.

\begin{figure}[tb!]
\includegraphics[width=5.5cm,clip]{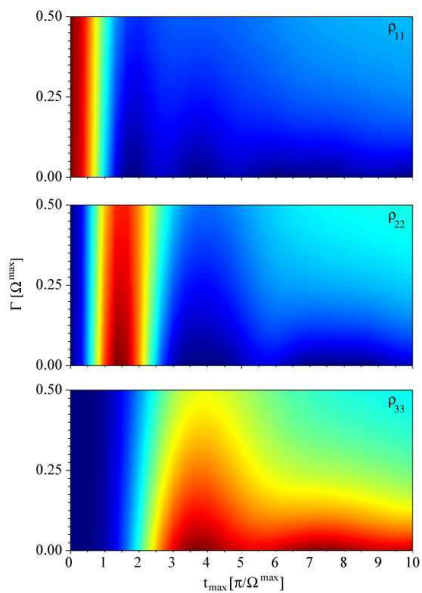}
\caption{\label{fig:CTAPME3d} Pseudo-color plots showing
populations (scale: red = 1, blue = 0) (a) $\rho_{11}$, (b) $\rho_{22}$ and
(c) $\rho_{33}$ as a function of total pulse time (in
units of $\pi/\Omega^{\max}$) and dephasing, $\Gamma$
(in units of $\Omega^{\max}$).  Successful transfer is
identified by $\rho_{11} \approx \rho_{22} \approx 0$ and
$\rho_{33} \approx 1$, this condition is satisfied when
$\Gamma/\Omega^{\max} \lesssim 0.1$, and
$t_{\max} \Omega^{max}/\pi \approx 3.75$.}
\end{figure}

We have considered the effects of $T_2$ dephasing in Fig.~\ref{fig:CTAPME3d} which shows the populations as a function of the total transfer time, $t_{\max}$ and $\Gamma$.  We can broadly identify three regimes here.  When $t_{\max} \Omega^{\max}/\pi \lesssim 2$, nonadiabatic transfer is observed, typified by the oscillating populations in $|1\rangle$ and $|2\rangle$.  Note that no appreciable population builds up in $|3\rangle$ in this regime.  For $t_{\max} \Omega^{\max}/\pi \gtrsim 5$ and $\Gamma/\Omega^{\max} \gtrsim 0.1$ the populations of all three states tend to equilibrate, again this regime is not useful for high fidelity transfer.  In order to maintain high fidelity, adiabatic transfer, we need $t_{\max} \Omega^{\max}/\pi \approx 3.75$ and $\Gamma/\Omega^{\max} \lesssim 0.1$.

That there is a window of transfer times to achieve high fidelity transfer should not be surprising, considering that the inclusion of $T_2$ effects introduces a characteristic timescale, beyond which coherence cannot be maintained.  So one must perform a transfer slow enough to be adiabatic, but fast enough to not be decohered.  Although these values for $\Gamma/\Omega^{\max}$ may appear small compared with optical systems, one should realise that in STIRAP we are comparing the transfer times with the \emph{two-photon} decoherence of the two ground states, which is normally very much longer than the decoherence rates of the optical transitions.  In our case we have assumed that the action of the environment is to decohere all superposition states equally as a pure dephasing process.  Pure dephasing is known to limit the performance of charge qubits and is manifested by electrostatic interactions with background charge fluctuators and phonon interactions \cite{bib:BarrettPRB2003,bib:FedichkinPRA2004}.

Because CTAP is a \emph{coherent} transfer method, there will be no heat dissipated by the electron during the transfer, and because the initial and final states are ground states, there can be no heat dissipation after transfer either.  Therefore, provided the transfer is fast relative to the decoherence time, there will be no heat dissipation \textit{regardless of the transfer speed}.  This should be contrasted with the quasi-adiabatic result for QDCA switching where there is a necessary tradeoff between switching time and heat dissipation \cite{bib:TimlerJAP2003}.  It must be stressed that although our scheme does not dissipate energy in the electronic system, there will of course be energy dissipation involved with charging/discharging the control gates.

\begin{figure}[tb!]
\includegraphics[width=6.5cm,clip]{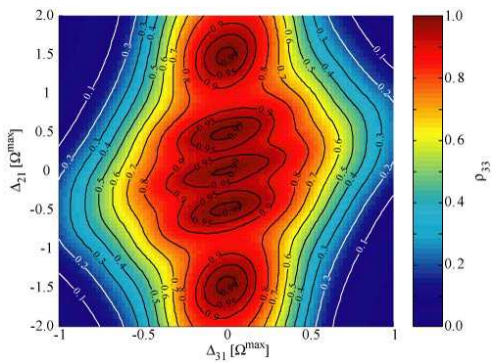}
\caption{\label{fig:VarE2E3} Pseudo-color plot with overlayed contours showing $\rho_{33}$ as a function of energy mismatches $\Delta_{31}$ and $\Delta_{21}$ in units of $\Omega^{\max}$ for $t_{\max} = 15 \pi /\Omega^{\max}$.  The system is less sensitive to $\Delta_{21}$ than $\Delta_{31}$ as expected for three-state adiabatic passage.}
\end{figure}

In the preceding discussion, we have assumed that the experimental parameters are known to arbitrary precision.  In general, however, the characterization of experimental Hamiltonians is a difficult problem, and complicated protocols must often be employed in order to extract the parameters to high fidelity (see for example Schirmer \textit{et al.}~\cite{bib:SchirmerPRA2004} for the case of a two-state system).  One advantage of STIRAP like schemes is that they are relatively insensitive to the exact conditions.  We have illustrated this in Fig.~\ref{fig:VarE2E3} where the effects of energy mismatches between the states have been modeled.  Here we show $\rho_{33}$ as a function of $\Delta_{31} = E_3 - E_1$ and $\Delta_{21} = E_2 - E_1$ for $t_{\max} = 15 \pi /\Omega$ to simulate the effect of a systematic error in the energy levels.  We have ignored the effects of dephasing for clarity in this figure. As is expected from the form of the dressed states in Eq.~\ref{eq:DressedStates} the behaviour is very insensitive to $\Delta_{21}$ but more strongly influenced by $\Delta_{31}$.  Note that an energy mismatch of as much as $\Omega^{\max}/5$ still permits $\rho_{33} \sim 0.9$.  One could imagine optimizing the transfer conditions by studying the $\rho_{33}$ and feeding results back to `tweak' the energies.

\section{Transport through a multi-dot system}
One obvious extension of this work is to consider the passage through more than one intervening dot.  This extension has been considered several times in the optical literature \cite{bib:VitanovARPC2001} and we concentrate here on one particular extension, the so-called straddling scheme of Malinovsky and Tannor \cite{bib:MalinovskyPRA1997}.  For simplicity we will not include the effects of dephasing in this treatment.

To realise the straddling tunnelling sequence (SCTAP) we must augment the original pulse sequence of Eq.~\ref{Eq:CTAPPulses} by the straddling pulses which are the same for all intervening tunnelling rates.  To be more precise, if we label the dots $1,2 \cdots n$ with tunnelling rates $\Omega_{12}, \Omega_{23} \cdots \Omega_{n-1,n}$ and we wish to transfer population from dot $1$ to dot $n$, then we apply the pulse sequence
\begin{eqnarray}
\Omega_{12}(t) \equiv \Omega_1 = \Omega_{CI}^{\max} \exp
    \left\{ -\frac{\left[ t- \left(t_{\max}/2+\sigma \right) \right]^2}
    {2 \sigma^2} \right\}, \nonumber \\
\Omega_{n-1,n}(t) \equiv \Omega_2 = \Omega_{CI}^{\max} \exp
    \left\{ - \frac{\left[t- \left(t_{\max}/2-\sigma\right) \right]^2}
    {2 \sigma^2} \right\}, \nonumber \\
\Omega_{i,i+1}(t) \equiv \Omega_S = \Omega_{S}^{\max} \exp
    \left[ - \frac{\left(t-t_{\max}/2 \right)^2}
    {4 \sigma^2} \right],
\label{Eq:SCTAPPulses}
\end{eqnarray}
where $1<i<n-1$, $\Omega_{CI}^{\max}$ is the maximum tunnelling rate for the counter-intuitive pulses, and $\Omega_{S}^{\max}$ is the maximum tunnelling rate for the `straddled' transitions.  Note that in contrast to the 3 state pulse sequence in Eq.~\ref{Eq:CTAPPulses} (with pulse separation $\sigma$), the spacing between pulses is $2\sigma$ for the straddling scheme.  The pulse separation in each case was chosen to optimize the transfer fidelity.  We are unable to provide a simple physical explanation for why the pulses separations in each case are different.

The Hamiltonian for the straddling scheme is ($\Delta=0$)
\begin{eqnarray}
\mathcal{H} = \left[
\begin{array} {ccccccc}
  0 & \Omega_{1} & 0 & \cdots & 0 & 0 & 0\\
  \Omega_{1} & 0 & \Omega_S &  & 0 & 0 & 0\\
  0 & \Omega_S & 0 & & 0 & 0 & 0\\
  \vdots & & & \ddots & & & \vdots\\
  0 & 0 & 0 & & 0 &\Omega_S & 0 \\
  0 & 0 & 0 & & \Omega_S & 0 &\Omega_{2} \\
  0 & 0 & 0 & \hdots & 0 & \Omega_{2} & 0 \\
\end{array}
\right].
\label{eq:HamSCTAP}
\end{eqnarray}
In contrast to Malinovsky and Tannor \cite{bib:MalinovskyPRA1997}, we find that this scheme is only effective for odd numbers of dots.  This discrepancy between even and odd numbers of states can be understood by examining the eigenvalue structure through an adiabatic passage.

\begin{figure}[tb!]
\includegraphics[width=5.5cm,clip]{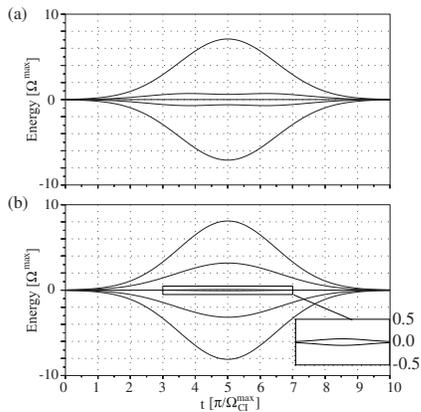}
\caption{\label{fig:EV56} Energies of the 5 (a) and 6 (b) dot systems with pulse sequences defined by Eq.~\ref{Eq:SCTAPPulses} and $t_{\max}=10 \pi /\Omega_{CI}^{\max}$ and $\Omega_{S}^{\max}=5 \Omega_{CI}^{\max}$.  Notice that the 6-state system has, in fact, two central states (with neither at zero energy) whereas the 5-state system has one central state at zero energy.  The two central states of the 6-state system are highlighted in the inset, which is an expanded view of the boxed region about the centre (the horizontal scale has been omitted for clarity, and is $[3-7]\pi/\Omega_{CI}^{\max}$).}
\end{figure}

In Fig.~\ref{fig:EV56} we show the eigenvalues for the 5 and 6 state SCTAP pulses with $\Omega_{S}^{\max} = 5 \Omega_{CI}^{\max}$ and $t_{\max} = 10 \pi /\Omega_{CI}^{\max}$.  Note that for 5 state transfer, there is one state at zero energy, with the other states symmetric about it, whereas for the 6 state system, there are two states close to zero.  These patterns are typical of all even and odd state systems.  The proximity of these two states to zero means that the adiabaticity criterion is much harder to satisfy for states with an even, rather than odd number of states.

The eigenvalues of the Hamiltonian in Eq.~\ref{eq:HamSCTAP} are in general rather complicated, however, the zero energy eigenstate has a relatively simple form in this case which is (unnormalized)
\begin{eqnarray}
|\mathcal{D}_0^{2n-1}\rangle &=& \cos \Theta_1 |1\rangle  - (-1)^{n} \sin \Theta_1 |2n-1\rangle \nonumber \\
&-& X \left[ \sum_{j=2}^{n-1} \left(-1\right)^j |2j-1\rangle \right],
\end{eqnarray}
where $\Theta_1$ is defined as above, and
\begin{eqnarray}
X = \frac{\Omega_{1}\Omega_{2}}{\Omega_{S} \sqrt{\Omega_{1}^2 + \Omega_{2}^2}}.
\end{eqnarray}
for an array of $2n-1$ dots.  We can now see that the scheme prevents occupation of the even numbered states, and minimizes population in the other undesirable states, as long as $X \ll 1$.  Note the similarities between these states, and the ones derived in the context of the chain $\Lambda$ scheme in Ref.~\cite{bib:GreentreePRA2003} which had alternating couplings and energy detunings.

\begin{figure}[tb!]
\includegraphics[width=5.5cm,clip]{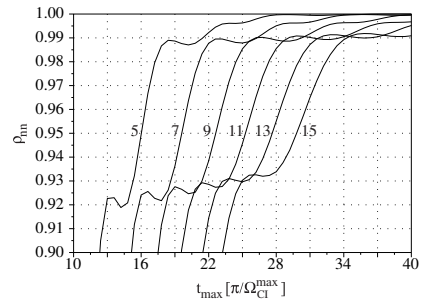}
\caption{\label{fig:HiFi579} $\rho_{nn}$ as a function of $t_{\max}$ (in units of $\pi/\Omega_{CI}^{\max}$) for (left to right along $t_{max}$ axis) 5,7,9,11,13, and 15 state systems for constant $\Omega_{S}^{\max} = 10 \Omega_{CI}^{\max}$.}
\end{figure}

To determine the parameter range needed to achieve high fidelity transfer, we solve the density matrix equations of motion for the multi-dot system, with varying $t_{\max}$ and $\Omega_{CI}^{\max}$ for constant $\Omega_{S}^{\max}=50$.  In Fig.~\ref{fig:HiFi579} we present results showing $\rho_{nn}$ as a function of the total transfer time, $t_{\max}$ for 5,7,9,11,13, and 15 state systems.  Note the almost monotonic behaviour with occasional `ripples'.  Because of the different pulse separations of the 3 state case, it is not valid to directly compare it with the other cases, although population transfer for the 3 state case is always easier to achieve than for other cases.

\section{Quantum information transfer across a linear chain}
We now consider applying the CTAP scheme to the problem of transferring quantum information across the chain.  This may be of use in quantum computation, where quantum information is required at a different point from where it is generated (e.g. taking information from a `computation zone' to a `measurement zone').  It is important to realise that the CTAP schemes as described are not capable of realizing swap operations without some modification, perhaps along the lines suggested by Unanyan \textit{et al.} \cite{bib:UnanyanPRA1999}.  However it \emph{is} possible to perform a transfer of quantum information across a network.  We consider the problem of attempting to switch the superposition state of a single charge-qubit, where the qubit is defined by a single electron in a superposition of two sites, $|\psi\rangle = \alpha|1\rangle + \beta|2\rangle$ with unoccupied sites $|3\rangle$ and $|4\rangle$.  We wish to transfer the complete quantum state in $|2\rangle$ to $|4\rangle$ leaving the rest of qubit unchanged.  Such information transfer would correspond to the following transition:
\begin{eqnarray}
\alpha|1\rangle + \beta|2\rangle \rightarrow \alpha|1\rangle + \beta|4\rangle.
\end{eqnarray}
One can show quite easily that such a transformation can be achieved by simply applying the standard CTAP pulse sequence to the $2,3,4$ system, without affecting state $|1\rangle$.  The Hamiltonian of the operation is
\begin{eqnarray}
\mathcal{H} = \left[
\begin{array}{cccc}
0 & 0 & 0 & 0 \\
0 & 0 & \Omega_1 & 0 \\
0 & \Omega_1 & 0 & \Omega_2 \\
0 & 0 & \Omega_2 & 0 \\
\end{array}
\right]
\end{eqnarray}
with $\Omega_1$ and $\Omega_2$ defined by $\Omega_{12}$ and $\Omega_{23}$ from Eq.~\ref{Eq:CTAPPulses} respectively.  If we start with our system in an arbitrary superposition state
\begin{eqnarray}
|\psi_i\rangle = \alpha|1\rangle + \beta|2\rangle
\end{eqnarray}
Numerically solving the master equations along the CTAP pulse trajectories yields the final state
\begin{eqnarray}
|\psi_f\rangle = \alpha|1\rangle + e^{i\pi}\beta|4\rangle
\end{eqnarray}
which is the desired final state up to a constant $\pi$ phase rotation, which can be understood by inspection of the eigenstates in Eq.~\ref{eq:DressedStates}.  We note that without maintaining the zero energy condition of the passage state (in this case the equivalent of the $|\mathcal{D}_0\rangle$ state from Eq.~\ref{eq:Eigenenergies}), the system would pick up an extra phase term due to the energy difference between state $|1\rangle$ and the adiabatically transformed state which depends critically on the \emph{exact} trajectory taken.  Such a consideration clearly shows the importance of adiabatic transform mechanisms where there is no change in energy and are therefore more robust against errors and fluctuations.

We may also consider the case where we wish to transfer both parts of the qubit using CTAP pulses.  Consider first the chain $|L3\rangle$, $|L2\rangle$, $|L1\rangle$, $|R1\rangle$, $|R2\rangle$, $|R3\rangle$ where we desire to transfer an arbitrary superposition of $|L1\rangle$ and $|R1\rangle$ to $|L3\rangle$ and $|R3\rangle$.  In this case we can apply the CTAP pulses simultaneously.  The Hamiltonian in this case is
\begin{eqnarray}
\mathcal{H} = \left[
\begin{array}{cccccc}
0 & \Omega_2 & 0 & 0 & 0 & 0 \\
\Omega_2 & 0 & \Omega_1 & 0 & 0 & 0\\
0 & \Omega_1 & 0 & 0 & 0 & 0 \\
0 & 0 & 0 & 0 & \Omega_1 & 0 \\
0 & 0 & 0 & \Omega_1 & 0 & \Omega_2 \\
0 & 0 & 0 & 0 & \Omega_2 & 0 \\
\end{array}
\right],
\end{eqnarray}
and the initial and final states are
\begin{eqnarray}
|\psi_i\rangle = \alpha|L_1\rangle + \beta|R_1\rangle, \nonumber \\
|\psi_f\rangle = \alpha|L_3\rangle + \beta|R_3\rangle,
\end{eqnarray}
respectively.  Note that in this case the $\pi$ rotations observed earlier have effectively cancelled so that the final state is in phase with the initial state.

One may extend these calculations to transfer across many dots using
an extension of the SCTAP sequence.  However for quantum information
transfer, transfer across an even number of dots will be
unsatisfactory as all states change their energy along the SCTAP
trajectory.

\section{Conclusions}

We have described a method of coherent electronic transport through
a triple-well system which we term CTAP (Coherent Tunnelling
Adiabatic Passage) by analogy with the optical process STIRAP
(Stimulated Raman Adiabatic Passage).  This scheme is realized by
electronic modulation coherent tunnelling rates by control of
barrier gates.  It is a high fidelity process, provided the coherent
tunnelling rates are fast compared with the T$_2$ dephasing times.
This scheme involves no change in the electronic energy throughout
the transfer regardless of transfer speed, and so provides an
interesting alternative to more conventional methods for
quasi-adiabatic clocking in QDCA cells.

The authors would like to thank G.~L.~Snider (Notre Dame), F.~Green (University of New South Wales), C.~J.~Wellard and S.~Prawer (University of Melbourne), and D.~K.~L.~Oi (University of Cambridge) for useful
discussions, and J. Mompart (Universitat Aut\`{o}noma de Barcelona) for making Ref.~\cite{bib:EckertPRA2004} known to us. This work was supported by the Australian Research
Council, the Australian government and by the US National Security
Agency (NSA), Advanced Research and Development Activity (ARDA)
and the Army Research Office (ARO) under contract number
DAAD19-01-1-0653.

%%%%%%%%%%%%%%%%%%%%%%%%%%%%%%%%%%%%%%%%%%%%%%%%%%%%%%%%%%%%%%%%

\end{document}